\documentclass[twocolumn]{aastex631}

\usepackage{amssymb}
\usepackage{array}
\usepackage{multirow}
\usepackage{bold-extra}
\AtBeginDocument{
}

\newcommand{\n}{\nodata}

\newcommand{\totalAGNs}{17478}
\newcommand{\jetdirAGNs}{9220}
\newcommand{\lastObsYear}{2021}
\newcommand{\totalImages}{123456}
\newcommand{\totalAGNsManyTimes}{4671}
\newcommand{\howManyTimes}{five}
\newcommand{\totalAGNsMultiFreq}{14735}
\newcommand{\totalAGNsCompleteSample}{3415}
\newcommand{\fracAtMainFreqs}{95\%}
\newcommand{\medianRedshift}{1.1}
\newcommand{\jetdirImages}{60594}
\newcommand{\countFlippedLowFreq}{200}
\newcommand{\fractionFlippedHighFreq}{0.5\%}

\received{2021 June 7}
\revised{2022 Mar 24}
\accepted{2022 Mar 25}

\submitjournal{ApJS}

\shorttitle{Parsec-Scale AGN Jet Directions}
\shortauthors{Plavin et al.}
\graphicspath{{./}{figs/}}

\begin{document}

\title{Direction of Parsec-Scales Jets for \jetdirAGNs{} Active Galactic Nuclei}

\correspondingauthor{A.~V.~Plavin}
\email{alexander@plav.in}

\author[0000-0003-2914-8554]{A.~V.~Plavin}
\affiliation{Lebedev Physical Institute of the Russian Academy of Sciences, Leninsky prospekt 53, 119991 Moscow, Russia}

\author[0000-0001-9303-3263]{Y.~Y.~Kovalev}
\affiliation{Lebedev Physical Institute of the Russian Academy of Sciences, Leninsky prospekt 53, 119991 Moscow, Russia}
\affiliation{Moscow Institute of Physics and Technology, Institutsky per. 9, Dolgoprudny 141700, Russia}
\affiliation{Max-Planck-Institut f\"ur Radioastronomie, Auf dem H\"ugel 69, 53121 Bonn, Germany}

\author[0000-0002-9702-2307]{A.~B.~Pushkarev}
\affiliation{Crimean Astrophysical Observatory, Nauchny 298409, Crimea, Russia}
\affiliation{Lebedev Physical Institute of the Russian Academy of Sciences, Leninsky prospekt 53, 119991 Moscow, Russia}

\begin{abstract}
The direction of parsec-scale jets in active galactic nuclei (AGNs) is essential information for many astrophysical and astrometric studies, including linear polarization and magnetic field structure, frequency-dependent synchrotron opacity, proper motion, and reference frame alignment.
We developed a rigorous, simple, and completely automated method to measure the directions from calibrated interferometric visibility data at frequencies ranging from 1.4~GHz to 86~GHz.
We publish the results for \jetdirAGNs{} AGNs with the typical accuracy below 10 degrees.
An internal check of the method comparing the directions between different observing frequencies as well as with previous publications verifies the robustness of the measured values.
\end{abstract}

\section{Introduction}
\label{s:intro}

Relativistic jets in active galaxies are prominent in a wide range of the electromagnetic spectrum. Radio emission is predominantly produced by the synchrotron mechanism and may span from sub-parsec to kiloparsec scales \citep{Blandford1979RelativisticJets}. The jet direction defines a distinguished axis of the AGN as a whole. This axis serves as a useful reference for other directional measurements: the polarization angle \citep[e.g.,][]{1985AJ.....90...30R,Kovalev2020OpticalPolarization}, apparent positional offsets between radio frequencies \citep[e.g.,][]{2011A&A...532A..38S,2012A&A...545A.113P,2018ARep...62..787V,2019MNRAS.485.1822P,
2020MNRAS.499.4515P} and between distant bands of the electromagnetic spectrum \citep[e.g.,][]{Kovalev2017VLBIGaia,2019ApJ...871..143P,Lambert2021}. Studying jet orientations also paves a way to astrophysical insights which include testing binary black hole models \citep[e.g.,][]{2012MNRAS.421.1861V},
evaluating properties of relativistic jet ejection \citep[e.g.,][]{2007A&A...476L..17A,2012ApJ...747...63A,2013AJ....146..120L}, and looking for the large-scale structure of the Universe \citep{2020A&A...635A.102B}.
Improving the current accuracy of absolute astrometry based on AGNs also requires taking the jet emission into account. This is particularly important when comparing reference systems obtained at different frequency bands \citep[e.g.,][]{2017MNRAS.471.3775P,2019MNRAS.482.3023P,2019ApJS..242....5X,2020A&ARv..28....6R,2020A&A...644A.159C} or studying AGN proper motion \citep[e.g.,][]{2011AJ....141..178M,2011A&A...529A..91T,2019MNRAS.482.3023P}.

AGN jets on parsec scales are highly collimated \citep[e.g.,][]{2017MNRAS.468.4992P,2020MNRAS.495.3576K}, but they still may bend due to interactions with external matter or other processes \citep{Gabuzda2001UnusualRadio,Kosogorov2021ParsecScale}. The brightest visible parsec-scale jets are hosted by blazars and have small viewing angles \citep{2017MNRAS.468.4992P}. Thus, we see them almost heads-on; such alignment makes these jets brighter due to relativistic beaming \citep{2007ApJ...658..232C,Kellermann2007DopplerBoosting,2019ApJ...874...43L}. These geometrical conditions lead to a strong amplification in the plane of the sky of any intrinsic changes in jet orientation. Thus, it is crucial to measure jet directions on the scales corresponding to the processes and regions of interest of each particular study. Many astrophysical studies, such as those listed above, focus on parsec or sub-parsec scales.

Very long baseline radio interferometry (VLBI) observations are well-suited for estimating parsec-scale jet directions. This is the only technique that directly resolves central parsecs for thousands of AGNs. Active galaxies have been observed with VLBI for several decades, leading to many discoveries, which are supplemented by images and raw observational data. These images have been used to estimate jet directions in multiple works already; see the references above. However, we are not aware of any available catalog with parsec-scale jet orientations systematically measured for thousands of objects. Creating such a catalog is the main goal of this work.

This paper presents a novel, completely automatic approach to measuring jet directions at (sub-)parsec scales based on VLBI observations. \autoref{s:obsdata} describes the source sample and observations used in this work. In \autoref{s:method}, we present our method, then in \autoref{s:evaluation}, we apply it to VLBI observations and compare it with previous results: \citet{2019ApJ...874...43L,2019ApJ...871..143P,2020A&A...635A.102B}. Jet directions for \jetdirAGNs{} objects are provided as electronic tables to support their usage by the community.

\section{Observational Data}
\label{s:obsdata}

For our analysis, we use VLBI observations at frequencies ranging from 1.4~GHz to 86~GHz compiled in the Astrogeo database\footnote{\url{ http://astrogeo.org/vlbi_images/}}, a collection from July~2021. It contains both the restored images and the visibility function measurements that are the original interferometric observables. We only rely on the latter in our analysis, and images are used solely for visualization. The database consists of geodetic VLBI observations \citep{2009JGeod..83..859P,2012A&A...544A..34P,2012ApJ...758...84P}, the VLBA\footnote{Very Long Baseline Array of the National Radio Astronomy Observatory, Socorro, NM, USA} calibrator surveys (VCS; \citealt{2002ApJS..141...13B,2003AJ....126.2562F,2005AJ....129.1163P,2006AJ....131.1872P,2007AJ....133.1236K,2008AJ....136..580P}), and other VLBI observations, including the results of \cite{2007ApJ...658..203H,2008AJ....136..159L,2011AJ....142...35P,2011MNRAS.414.2528P,2011AJ....142..105P,2012MNRAS.419.1097P,2013AJ....146....5P,2015ApJS..217....4S,2017ApJS..230...13S,2017ApJ...846...98J,2018ApJS..234...12L,2019MNRAS.485...88P,2019A&A...622A..92N,2021AJ....161...14P,2021AJ....161...88P}.

The full dataset used for our analysis contains \totalAGNs{} sources observed from 1994 to \lastObsYear{}, \totalImages{} single frequency images in total. Out of those AGNs, \totalAGNsManyTimes{} were observed at least \howManyTimes{} times, and \totalAGNsMultiFreq{} at two or more frequency bands. 
We note that an all-sky complete flux-density-limited sample was constructed within the VCS program, and consists of \totalAGNsCompleteSample{} targets stronger than 150~mJy at 8~GHz.
The majority of observations, \fracAtMainFreqs{} of all source-epochs, are at 2~GHz, 5~GHz, 8~GHz, or 15~GHz. The observed objects are radio-bright active galaxies with VLBI flux densities ranging from a few millijanskys to tens of janskys. According to the NED, these AGNs have the median redshift of \medianRedshift{} and belong to various subclasses. Flat-spectrum radio quasars constitute about a third of the sample, BL Lac objects and radio galaxies make up 10\% each, and 3\% of Seyfert galaxies. The remaining 40\% of the sample do not have their optical class reliably determined; these sources are also somewhat weaker on average in terms of their VLBI flux. The sample is dominated by blazars, i.e. AGNs with a small viewing angle of a few degrees. See \citet{2019ApJ...874...43L} for a detailed study of a complete flux density-limited sample with the MOJAVE program at 15~GHz.

\begin{deluxetable*}{lRCRRrcRC}
\tablecaption{Measurements of the AGN jet directions separately for each frequency band. \label{t:jetdirs_perband}}
\tablehead{
\colhead{J2000 Name} & \colhead{Frequency} & \colhead{$N_\mathrm{epochs}$} & \colhead{PA} & \colhead{PA Error} & \colhead{PA Error Type} & \colhead{$180^\circ$} &\multicolumn{2}{c}{Core to Jet Distance} \\[-5pt]
\colhead{} & \colhead{(GHz)} & \colhead{}& \colhead{($^\circ$)} & \colhead{($^\circ$)} & \colhead{}& \colhead{Correction} & \colhead{median (mas)} & \colhead{MAD (mas)}
}
\decimalcolnumbers
\startdata
J1256$-$0547 & 1.4 &   1 & $-137$ &       4 &    ASTD &  -- &  5.3 & \nodata\\
J1256$-$0547 &   2 &   3 & $-123$ &       4 &    ASTD & F15 &  2.4 &     0.2\\
J1256$-$0547 &   5 &   1 & $-132$ &       5 &    ASTD &  -- &  6.8 & \nodata\\
J1256$-$0547 &   8 &   9 & $-116$ &      13 &     STD &  -- &  3.0 &     0.5\\
J1256$-$0547 &  15 & 146 & $-130$ &      13 &     STD &  -- &  0.9 &     0.3\\
J1256$-$0547 &  24 &   2 & $-143$ &       7 &    ASTD &  -- &  0.9 &     0.1\\
J1256$-$0547 &  43 & 116 & $-147$ &      10 &     STD &  -- &  0.7 &     0.2\\
J1256$-$0547 &  86 &   1 & $-142$ & \nodata &      -- &  -- &  0.4 & \nodata\\
  J2344+2952 &   2 &   5 &  $-93$ &       1 &     STD &  F8 & 12.5 &     0.2\\
  J2344+2952 &   8 &   4 &  $-74$ &       5 &    ASTD &  -- &  1.9 &     1.0
\enddata
\tablecomments{Columns are as follows:
(1)~--~J2000 name;
(2)~--~frequency band;
(3)~--~number of VLBI epochs contributing to the measurement;
(4)~--~jet position angle;
(5)~--~uncertainty of the position angle;
(6)~--~nature of the uncertainty estimate: intra-band standard deviation (`STD') if $N_\mathrm{epochs} \geq 5$; otherwise average standard deviation at this band (`ASTD') for frequencies below 86~GHz, and none for 86~GHz;
(7)~--~$180^\circ$ correction flag: based on the orientation that we measure at a higher frequency of $\nu$~GHz (`F$\nu$') or based on dedicated studies of specific objects, J0900$-$2808 in \citealt{Kosogorov2021ParsecScale} (`E1') and J0927+3902 in \citealt{1993ApJ...402..160A} (`E2');
(8)~--~the median distance between the core and the jet components, indicating angular scales probed by this measurement;
(9)~--~the median absolute deviation of the core-jet component distance when at least two epochs are available.
\\
The table is published in its entirety in the machine-readable format. A portion is shown here for guidance regarding its form and content.}
\end{deluxetable*}

\begin{deluxetable*}{lLRRcCC}
\tablecaption{Frequency-averaged directions of parsec-scale jets with a single value for each AGN. Contains the same objects as \autoref{t:jetdirs_perband}, and provides our final estimates of the jet orientation. This table should generally be used unless individual frequencies and corresponding spatial scales are specifically of interest. \label{t:jetdirs_agg}}
\tablehead{
\colhead{J2000 name} & \colhead{Frequencies} & \colhead{PA} & \colhead{PA~Error} & \colhead{$180^\circ$} & \multicolumn{2}{c}{Core to Jet Distance} \\[-5pt]
\colhead{} & \colhead{(GHz)} & \colhead{($^\circ$)} & \colhead{($^\circ$)} & \colhead{Correction} & \colhead{median (mas)} & \colhead{MAD (mas)}
}
\decimalcolnumbers
\startdata
J0509+0541 & 2, 5, 8, 15, 24 & $-177$ & 3 & -- & 2.0 &     0.9\\
J0823+2223 & 5, 8, 15 & $-164$ & 3 & -- & 4.7 &     0.6\\
J0927+3902 & 1.4, 2, 5, 8, 15, 24 &  100 & 3 & E2 & 1.6 &     0.6\\
J1256$-$0547 & 1.4, 2, 5, 8, 15, 24, 43, 86 & $-133$ & 4 & -- & 0.9 &     0.7\\
J2344+2952 & 2, 8 &  $-83$ & 3 & -- & 7.2 &     5.3
\enddata
\tablecomments{Columns are as follows:
(1)~--~J2000 name;
(2)~--~frequency bands contributing to this measurement;
(3)~--~jet position angle;
(4)~--~uncertainty of the position angle;
(5)~--~$180^\circ$ correction flag based on dedicated studies of specific objects: J0900$-$2808 in \citealt{Kosogorov2021ParsecScale} (E1) and J0927+3902 in \citealt{1993ApJ...402..160A} (E2);
(6)~--~the median distance between the core and the jet components, indicating angular scales probed by this measurement;
(7)~--~the median absolute deviation of the core-jet component distance when at least two frequencies are available.
\\
The table is published in its entirety in the machine-readable format. A portion is shown here for guidance regarding its form and content.}
\end{deluxetable*}

\section{Determining the Jet Direction}\label{s:method}

Studies that make use of parsec-scale jet orientations often estimate them from VLBI images. This was realized both in a fully automated way \citep[e.g.,][]{Kovalev2017VLBIGaia,2019ApJ...871..143P} and semi-automatically with manual intervention \citep[e.g.,][]{2020A&A...635A.102B}. Independently of the specific approach, these measurements are fundamentally limited by the information available in the images. Typical VLBI images cannot reach the maximum resolution in the brightest areas \citep[e.g.,][]{Hogbom1974ApertureSynthesis}. Estimation of the jet direction close to the origin is critically dependent on these central regions. Their blending may even make the measurement impossible when the jet brightness quickly falls further downstream.

An alternative is to determine the jet orientation without relying on images by directly analyzing the calibrated visibilities instead. Visibilities, the principal VLBI observables, correspond to the Fourier transform of the brightness distribution in the sky. They are commonly used to infer the brightness distribution itself via a process called ``model-fitting'': the observed structure is assumed to consist of a number of Gaussian components, whose parameters are fit to match the visibilities. This approach achieves a higher effective resolution in high-SNR regions compared to that of images and allows separating multiple bright features located close to each other. Model-fitting has been performed for VLBI observations in numerous works. Common limitations of model-fitting include the semi-manual nature of the method, requiring human-dependent decisions on the model being fitted. A notable massive application of this technique is regularly performed within the MOJAVE program \citep{2019ApJ...874...43L}, which costs hundreds of human hours not only producing models but also analyzing and discussing their robustness by the MOJAVE team. The detailed results of their model fitting are publicly available and can be used to determine jet directions \citep[e.g.,][]{2013AJ....146..120L} among other properties. However, these results cover only 15~GHz observations of 382 brightest jets in the Northern sky, a relatively small subsample of all AGNs observed with VLBI.
Automated visibility model-fitting was also performed by, e.g., \citet{2005AJ....130.2473K,2015MNRAS.452.4274P} in \texttt{difmap} \citep{Shepherd1994DIFMAPInteractive}. Application of its results was typically limited to parameters of the dominant component.

\begin{figure}
\centering
\includegraphics[height=0.3\textheight]{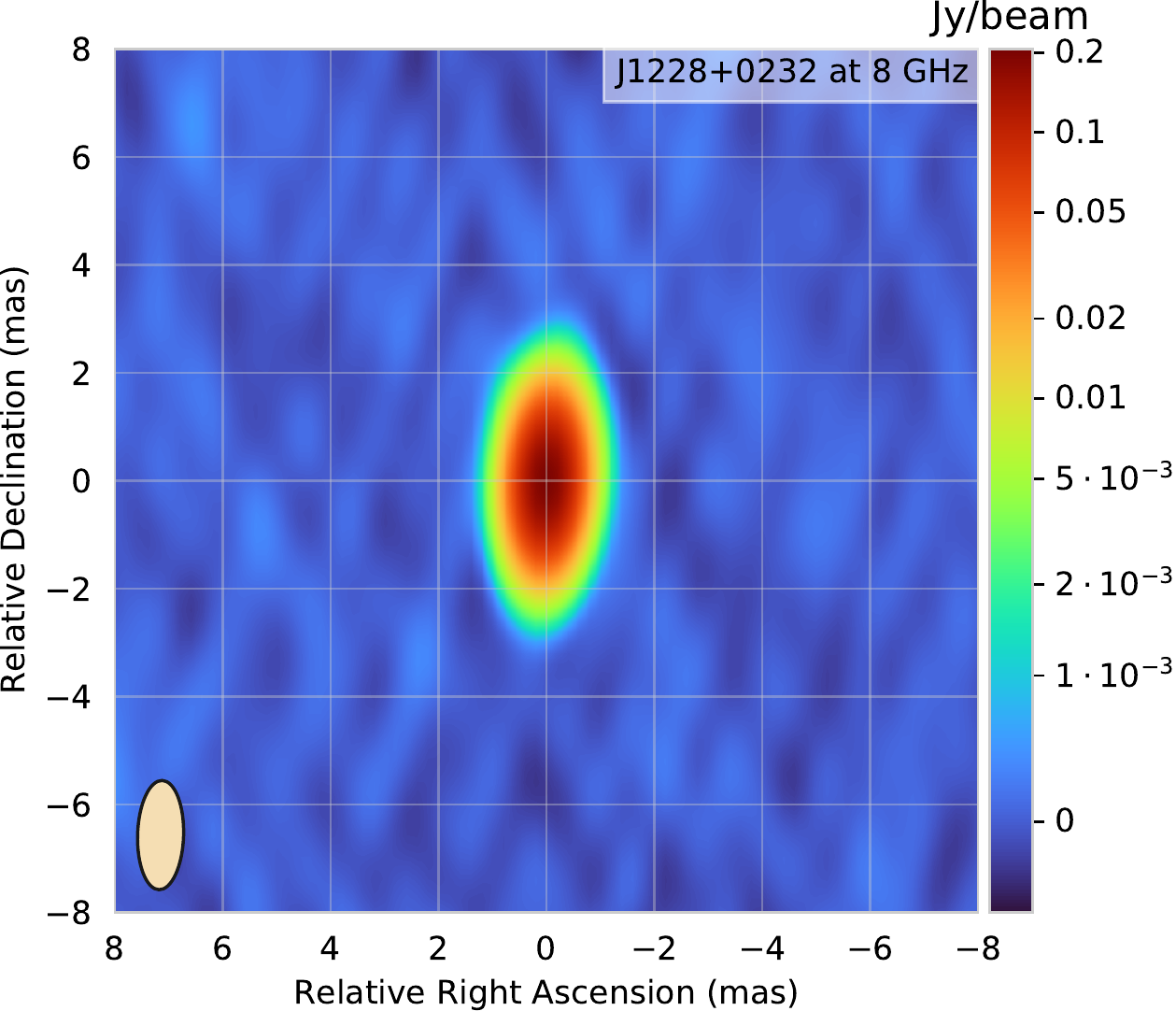}
\caption{Example of a VLBI total intensity map with no detected jet structure. Here and in the subsequent figures, the beam size (FWHM) is indicated by the ellipse in the lower left corner.}
\label{f:example_nojet}
\end{figure}

\begin{figure}
\centering
\includegraphics[height=0.3\textheight]{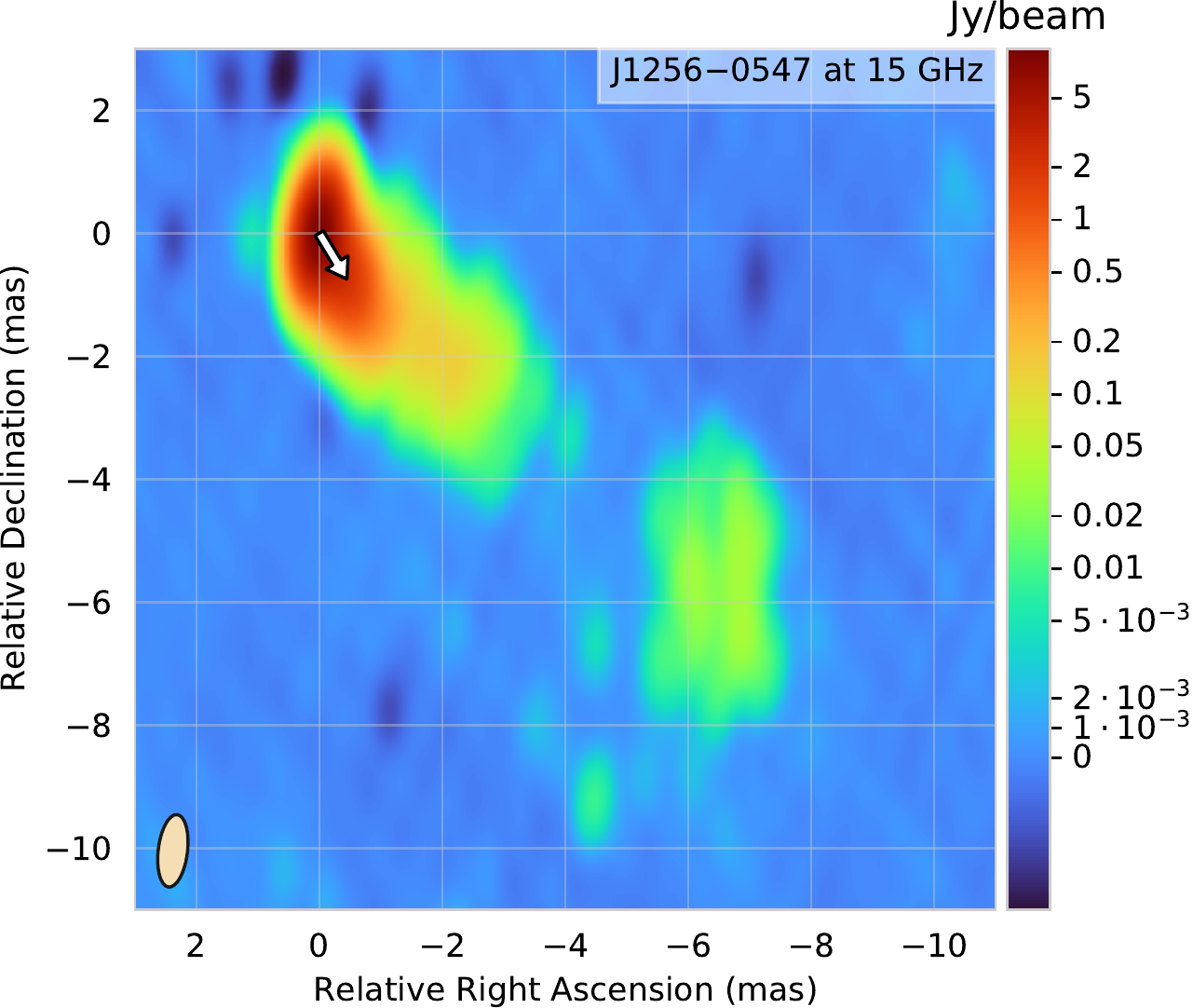}
\caption{Example of a VLBI image with an extended jet and a successfully measured jet direction indicated by the arrow. The arrow here and in the figures below indicates the jet orientation, as determined in our analysis. The arrow ends mark the positions of the two Gaussian components in our structure model.}
\label{f:example_common}
\end{figure}

We base our procedure of estimating jet directions on the model-fitting approach. Specifically, we fit two simple models to the calibrated visibilities of each source-epoch: one is a single circular Gaussian component, and the other is two Gaussian components. Fitting is performed via a Bayesian nested sampling algorithm as implemented in the \texttt{PolyChord} library \citep{2015MNRAS.450L..61H}. Compared to the maximum likelihood fitting implemented in the \texttt{difmap} package, this algorithm does not require an initial guess of parameter values and delivers more principled uncertainty estimates. Note that these uncertainties are still fundamentally underestimated, as they do not account for model assumptions and the calibration performed beforehand. Despite this limitation, we find formal uncertainties a useful proxy for actual errors, as discussed further in \autoref{s:evaluation}. For the purpose of jet direction measurement, we drop source-epochs with formal uncertainties exceeding $45\degr$. Additionally, nested sampling algorithms provide the so-called evidence that aids comparison between different models. We only keep source-epochs better described by the two-component model, judging by the evidence value: other observations effectively do not contain any detectable extended jet structure.

Based on the two-Gaussian model fitted to visibilities, we calculate the jet position angle as the direction from one component to the other. Formal uncertainties, which result from the fitting, are propagated through these calculations. It remains to determine which of the two components is closer to the apex and which is a downstream feature in the jet. Further, we call the bright compact component at the apparent jet origin \emph{the VLBI ``core''} \citep{1981ApJ...243..700K}. There is no objective and fundamentally correct way to choose the core component in general, so we develop a heuristic approach. Regions closer to the jet origin are most commonly brighter in terms of their intensity or effective temperature \citep[e.g.,][]{2005AJ....130.2473K,2006ApJ...642L.115H}. This effect has solid theoretical grounds \citep[e.g.,][]{Blandford1979RelativisticJets}. We thus propose to select the brightest component of the two as the apparent jet origin. This selection criterion crucially relies on adequate intensity measurements. They would be hard, if possible, to obtain in image-based methods, but are directly available from our visibility model fitting. We compute the intensity as $I = S / \theta^2$, where $S$ is the flux density, and $\theta$ is the Gaussian effective angular size. The brightest component is then selected unless $\theta$ is poorly determined and has a relative uncertainty above 50\%. In these cases, we chose the strongest component in terms of its flux density $S$ as the jet origin.

It is common to have multiple VLBI observations for an AGN at the same frequency performed on different epochs (\autoref{s:obsdata}). In these cases, we take a median of individual jet directions as the resulting measurement at this frequency, making the estimate more accurate. A by-product of this aggregation is an alternative uncertainty estimate of the jet orientation: namely, the standard deviation of individual directions. This is a conservative upper bound of the uncertainty and includes real changes with time. It is thus directly useful as a measure of how well each object is characterized with a single jet orientation estimate at a specific frequency. This is our focus in this work, while a detailed evaluation of the real jet orientation variability \citep[e.g.,][]{2013AJ....146..120L} is currently out of scope.

We combine jet direction measurements across all available frequencies. Multi-frequency observations allow us to resolve the $180^\circ$ jet direction ambiguity more reliably. The radio spectrum of the extended optically thin jet is typically steeper than that of the partially opaque core \citep[e.g.,][]{2014AJ....147..143H,2019MNRAS.485.1822P}. Thus, the observed structure is increasingly core-dominated at higher frequencies. A direct spectrum calculation is challenging because of different resolutions and is further complicated by the time variability. We follow a simplistic approach: whenever the estimated jet orientations at two frequencies differ by more than $90^\circ$, the one at the lower frequency is inverted, i.e., rotated by $180^\circ$. The resulting jet position angles are listed in \autoref{t:jetdirs_perband}, together with the flags indicating whether this inversion was applied.

Finally, we obtain a single direction estimate for each AGN by averaging the measurements at all individual frequencies. These averages are presented in \autoref{t:jetdirs_agg}. They are useful as the most aggregated direction estimates, despite being potentially affected by jet bending. We do not perform any weighting so that all frequencies contribute equally to the average. This may increase the statistical error when there are many more epochs in one band than in another. However, we always follow this approach to achieve treatment of all objects as uniform as possible.

\section{Method Evaluation and Comparison}\label{s:evaluation}

\begin{figure}
\centering
\includegraphics[width=\linewidth]{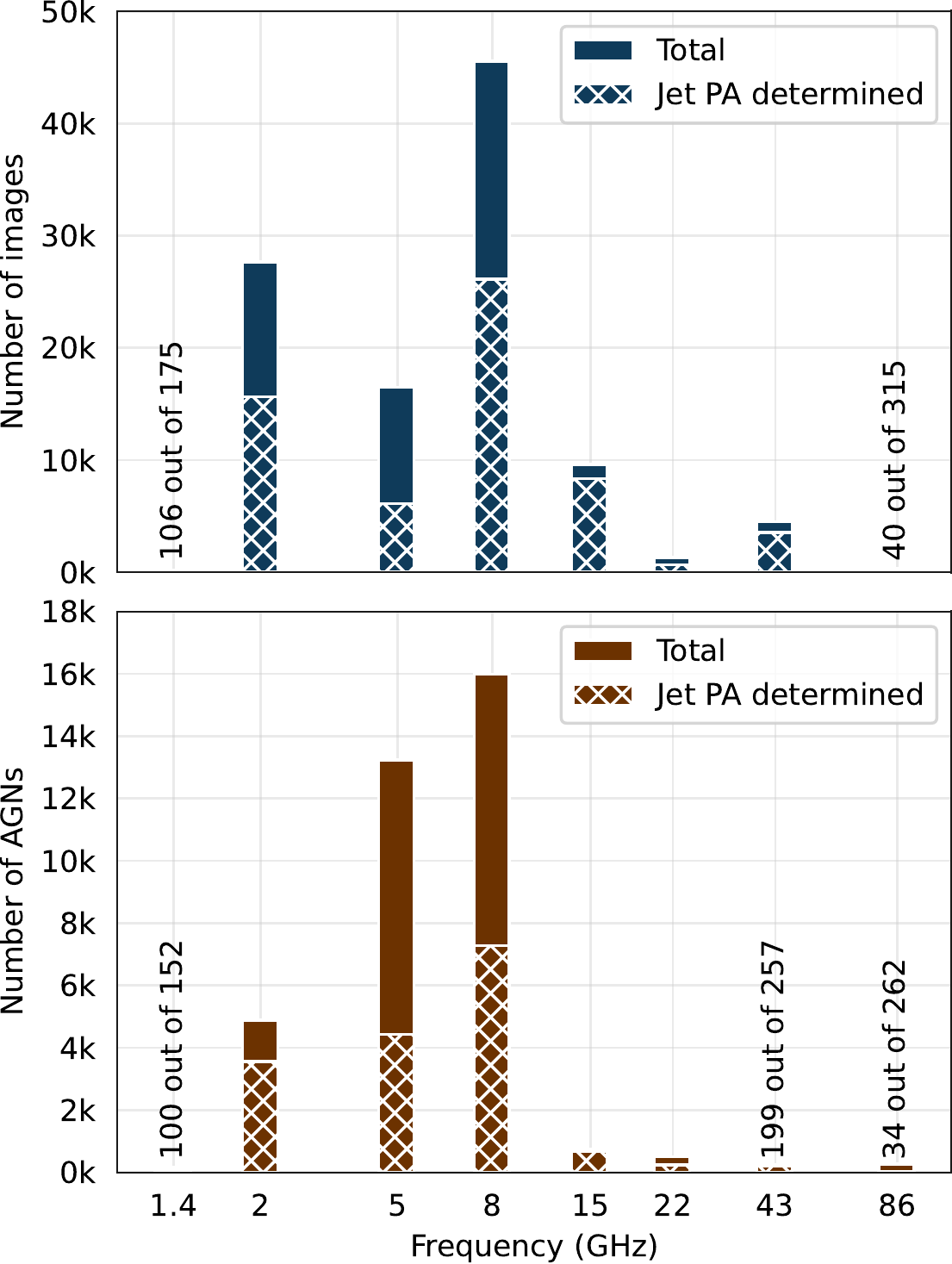}
\caption{Number of individual VLBI epochs and AGNs: all available in the Astrogeo database, and those where we managed to determine the jet direction.}
\label{f:images_cnt}
\end{figure}

In this section, we evaluate and characterize the achieved performance and the robustness of our jet direction measurement method, including comparison with previous studies.
First, many VLBI observations just do not reveal any reliably detected jet structure outside the core, and the resulting images look like \autoref{f:example_nojet}. We find them to generally be better described by a single-component model or yield large formal uncertainties of the jet direction (see \autoref{s:method}). Thus, we drop such observations from further analysis.
For comparison, an example of a typical image with a long extended jet is shown in \autoref{f:example_common}. After the filtering, \jetdirImages{} source epochs remain, which is {\the\numexpr100*\jetdirImages/\totalImages}\%  of the total sample; \autoref{f:images_cnt} shows a breakdown by the frequency band. All corresponding single-band jet orientation measurements are presented in \autoref{t:jetdirs_perband}.

\begin{figure}
\centering
\includegraphics[width=\linewidth]{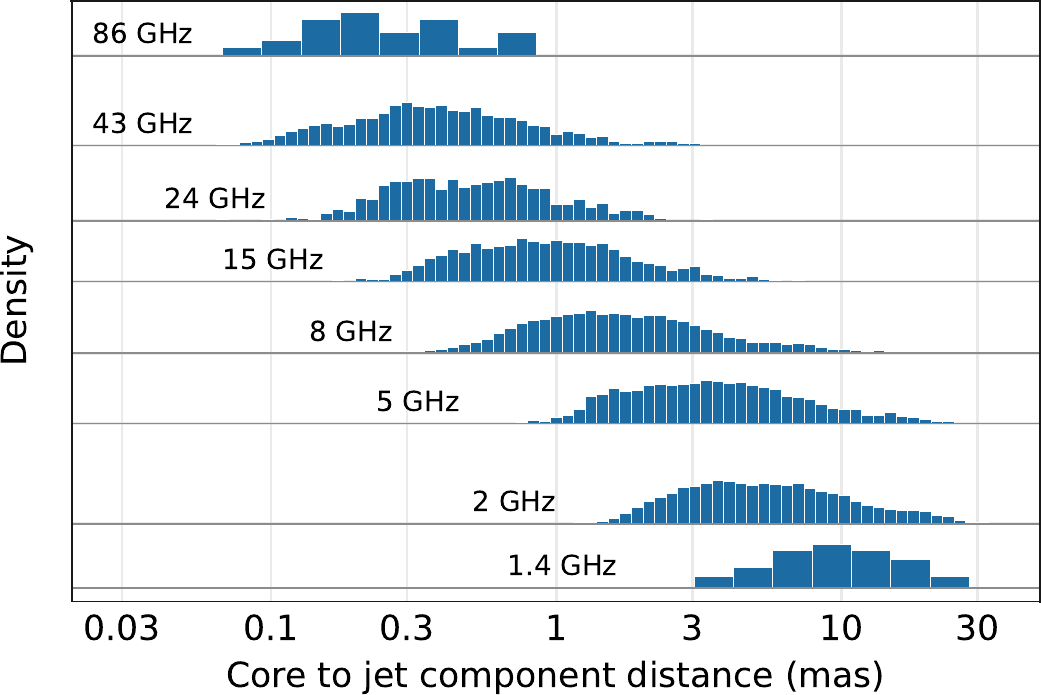}
\caption{Distribution of angular distances between the core and the jet components for jet direction measurements at each frequency band. These distances indicate the scales probed by our analysis.}
\label{f:corejet_dist}
\end{figure}

Observations at different frequencies are sensitive to and probe different angular scales in the jet due to resolution effects and intrinsic properties of synchrotron emission. Scales probed by our direction measurements correspond to distances between the core and the jet components. They vary from hundreds of microarcseconds to tens of milliarcseconds and are listed in tables alongside orientation estimates. Distributions of these distances are shown in \autoref{f:corejet_dist}.

Typical VLBI observations provide significantly non-uniform coverage of the visibility plane, especially in the North-South versus East-West directions. This irregularity results in two orthogonal directions of the best and worst angular resolution, commonly referred to as the beam minor and major axes. We attempt to quantify the systematic effects of this non-uniformity on our orientation measurements. The measured core to jet components distance, as defined above, is 20\% to 30\% larger on average along the beam major axis compared to the minor axis direction. At the same time, we do not detect any strong preference of jet directions to align with the beam: differences from the uniform distribution are below 10\% at frequencies from 2~GHz to 15~GHz and at 20\% or below for all other bands.

\begin{figure}
\centering
\includegraphics[width=1.0\linewidth]{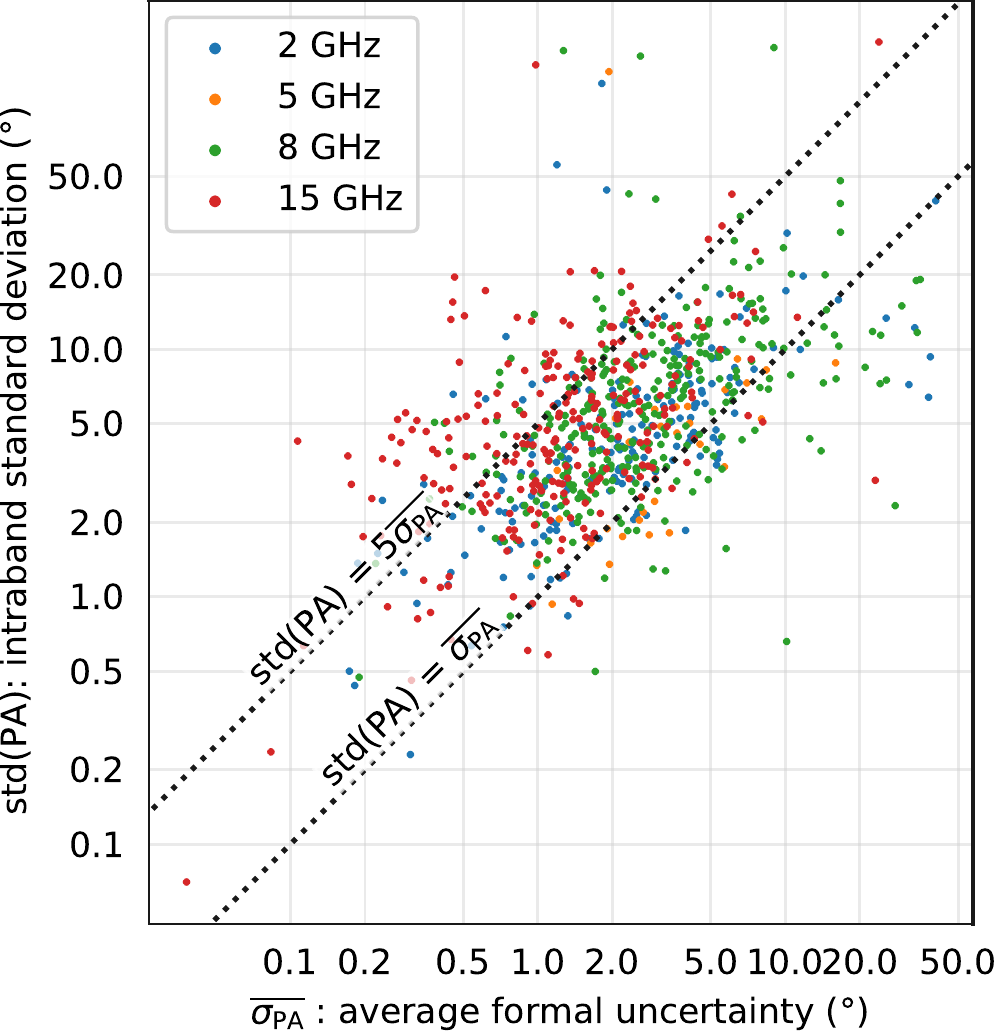}
\caption{Comparison of the average formal model-fitting uncertainty $\overline{\sigma_\mathrm{PA}}$ derived from individual single-epoch measurements with the intra-band scatter $\mathrm{std}(\mathrm{PA})$. Only AGNs with at least ten measurements at the corresponding frequency are shown here. The majority of the points lie below the top dotted line: the scatter is less than five times the formal uncertainty for about 90\% of the objects.}
\label{f:formal_vs_std}
\end{figure}

We consider statistical model-fitting uncertainties in the jet position angle as lower bounds of the true errors (\autoref{s:method}). It is thus instructive to compare these formal uncertainties to the scatter of the measured jet directions of a single object. \autoref{f:formal_vs_std} illustrates this comparison with the intra-band scatter quantified by the standard deviation. The distribution qualitatively looks as expected: the standard deviation is larger than the formal uncertainty in almost all cases, and a clear correlation is present. The uncertainties we provide in \autoref{t:jetdirs_perband} are based on this intra-band scatter. Specifically, we use the standard deviation for objects with at least five epochs at a single frequency available. The uncertainties given for AGNs with fewer epochs are averages of these deviations within each frequency band; they only reflect average properties at a given frequency. There are no sources with at least five direction measurements at 86~GHz, and uncertainties are not provided at this frequency. We do not directly utilize model-fitting errors to estimate the resulting uncertainties because formal errors are systematically lower than the intra-band scatter and thus would impede uniform comparisons between AGNs. The intra-band scatter forms a conservative uncertainty estimate because true orientation variations are possible \citep[e.g.,][]{2013AJ....146..120L}. Nevertheless, such estimates are useful when describing the geometry of an object with a single jet direction assessment.

\begin{figure*}
\centering
\includegraphics[height=0.3\textheight]{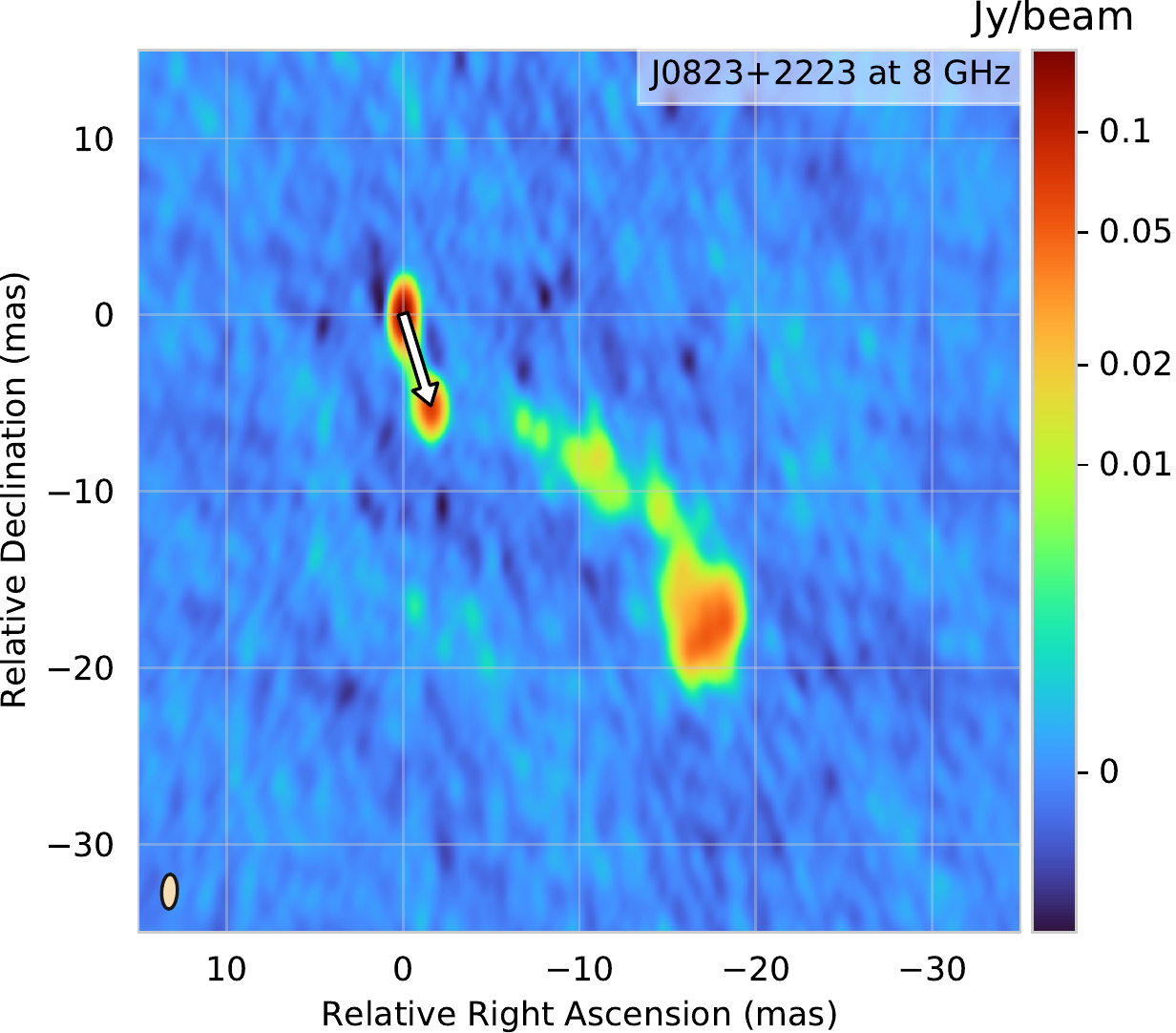}
\includegraphics[height=0.3\textheight]{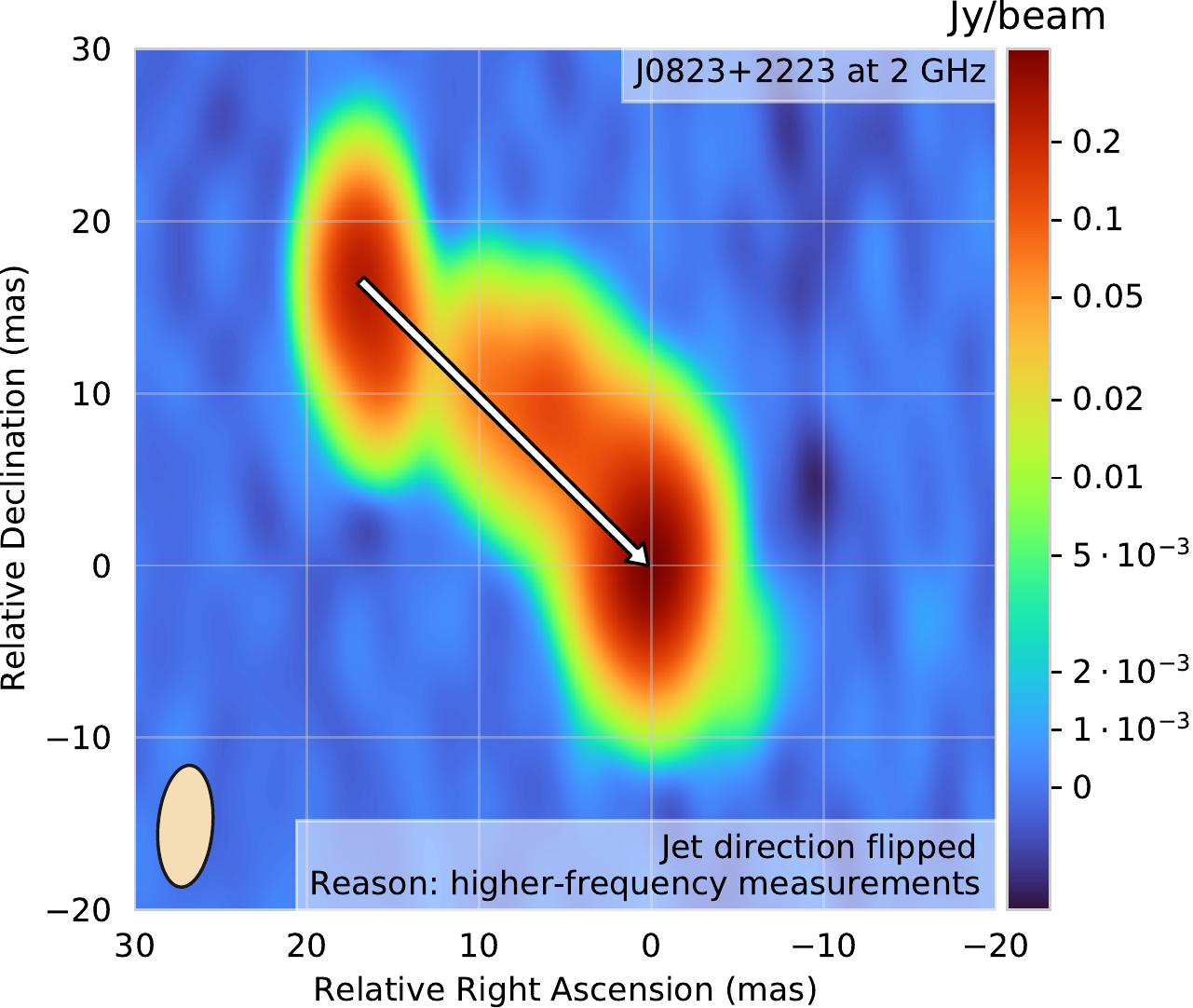}
\caption{VLBI images of the same AGN at 8 and 2 GHz. The lower-frequency jet direction was flipped following the criteria outlined in \autoref{s:method}. The residual difference between these directions is attributed to apparent jet bending.}
\label{f:example_diff28}
\end{figure*}

\begin{figure*}
\centering
\includegraphics[height=0.3\textheight]{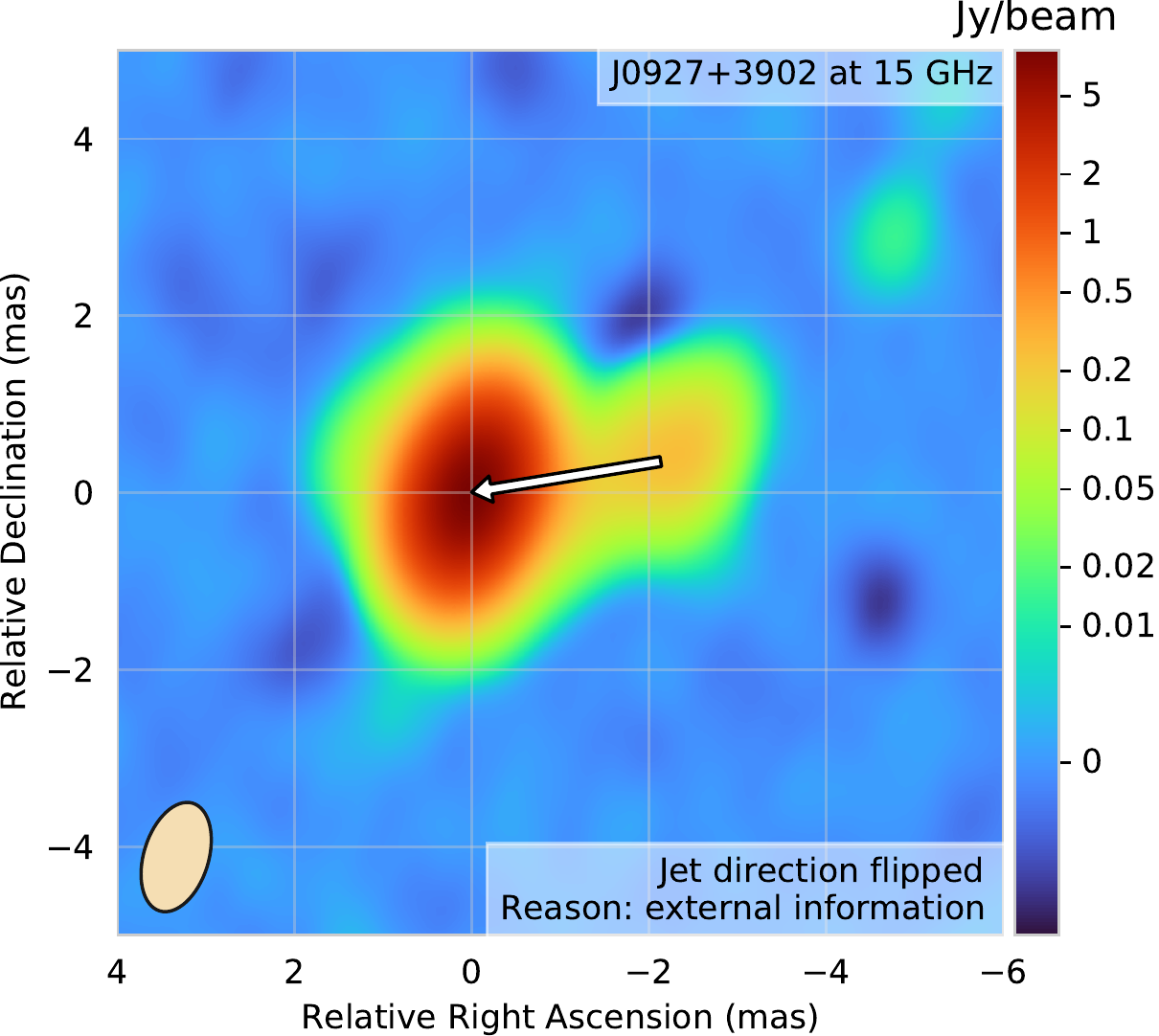}
\includegraphics[height=0.3\textheight]{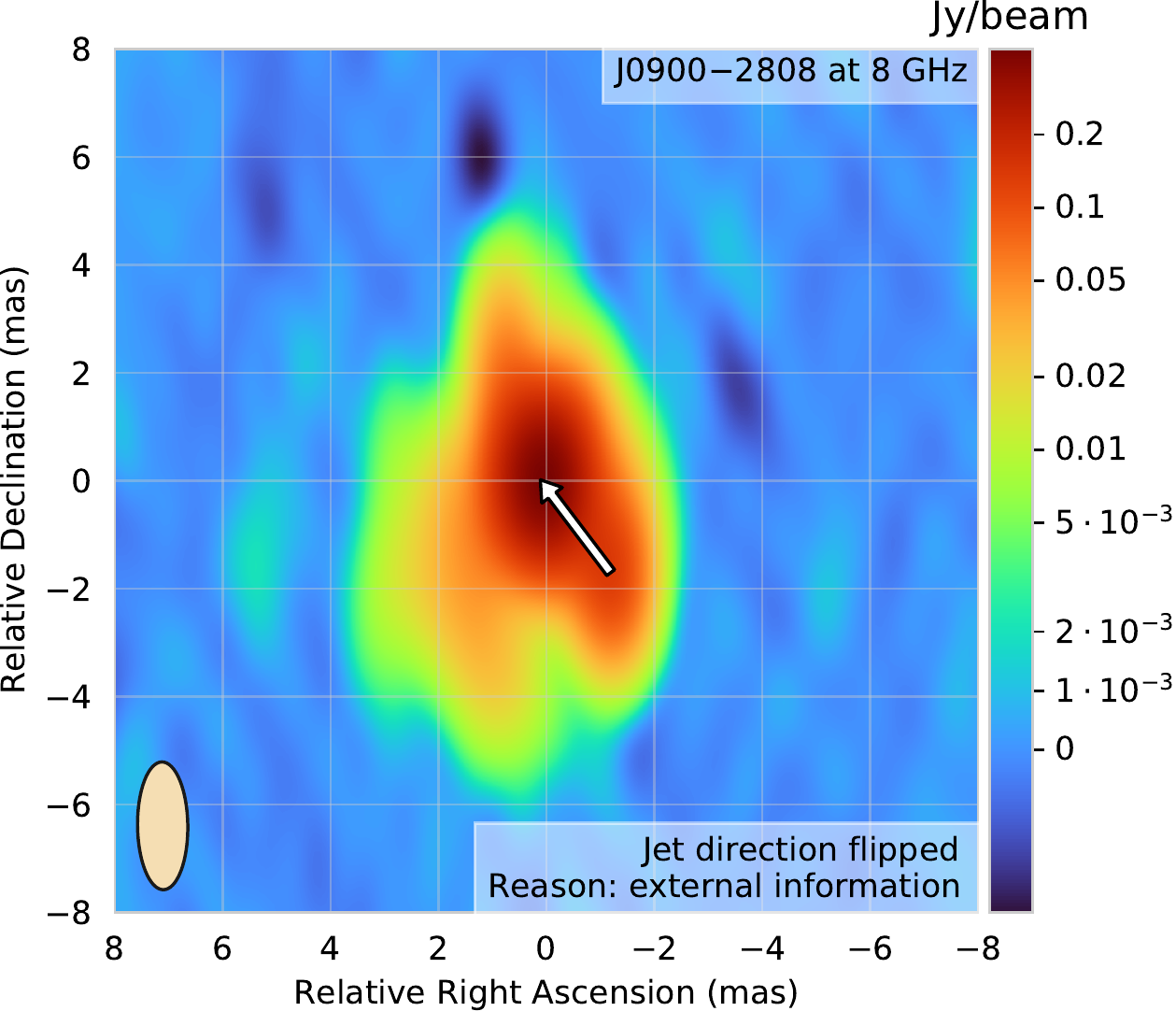}
\caption{VLBI images of the two AGNs where we had to explicitly flip the jet direction due to the available information from specific detailed studies. See the discussion in \autoref{s:evaluation}.}
\label{f:example_flipped}
\end{figure*}

\subsection{Multi-Frequency Effects}

As discussed in \autoref{s:method}, the extended jet regions typically have steeper spectra compared to the VLBI core. This can make jets stronger, easier to detect, and more extended at lower frequencies. At the same time, it becomes more probable to incorrectly resolve the $180^\circ$ ambiguity in their direction: the jet component may become stronger than the core, effect that was noted before in both astrophysical \citep{Kovalev2008OpacityCompact} and astrometrical \citep{Xu2021EvidenceGaia} contexts. An example in \autoref{f:example_diff28} shows a case when the jet direction differs between the frequencies. Note that in this example, the core component is brighter than the features in the jet at 8~GHz. However, due to the spectral properties it becomes fainter at 2~GHz. Following \autoref{s:method}, we flip the direction estimated at the lower frequency in such cases; the arrow in \autoref{f:example_diff28} illustrates the corrected direction. There are around \countFlippedLowFreq{} objects with their 2~GHz or 5~GHz jet orientation flipped following this criterion, and less than \fractionFlippedHighFreq{} at each of the higher frequencies.

\begin{figure*}
\centering
\includegraphics[height=0.22\textheight]{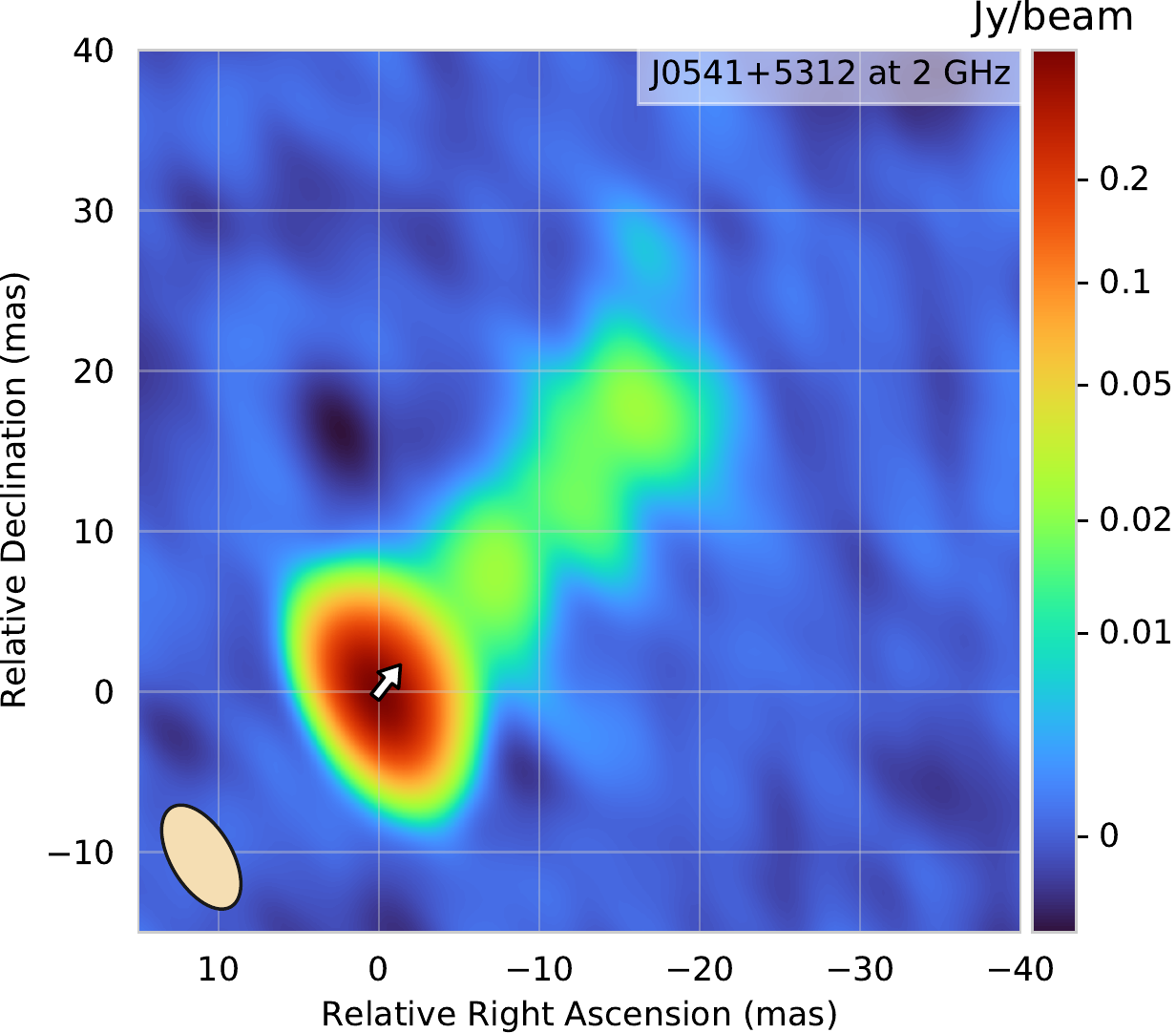}
\includegraphics[height=0.22\textheight]{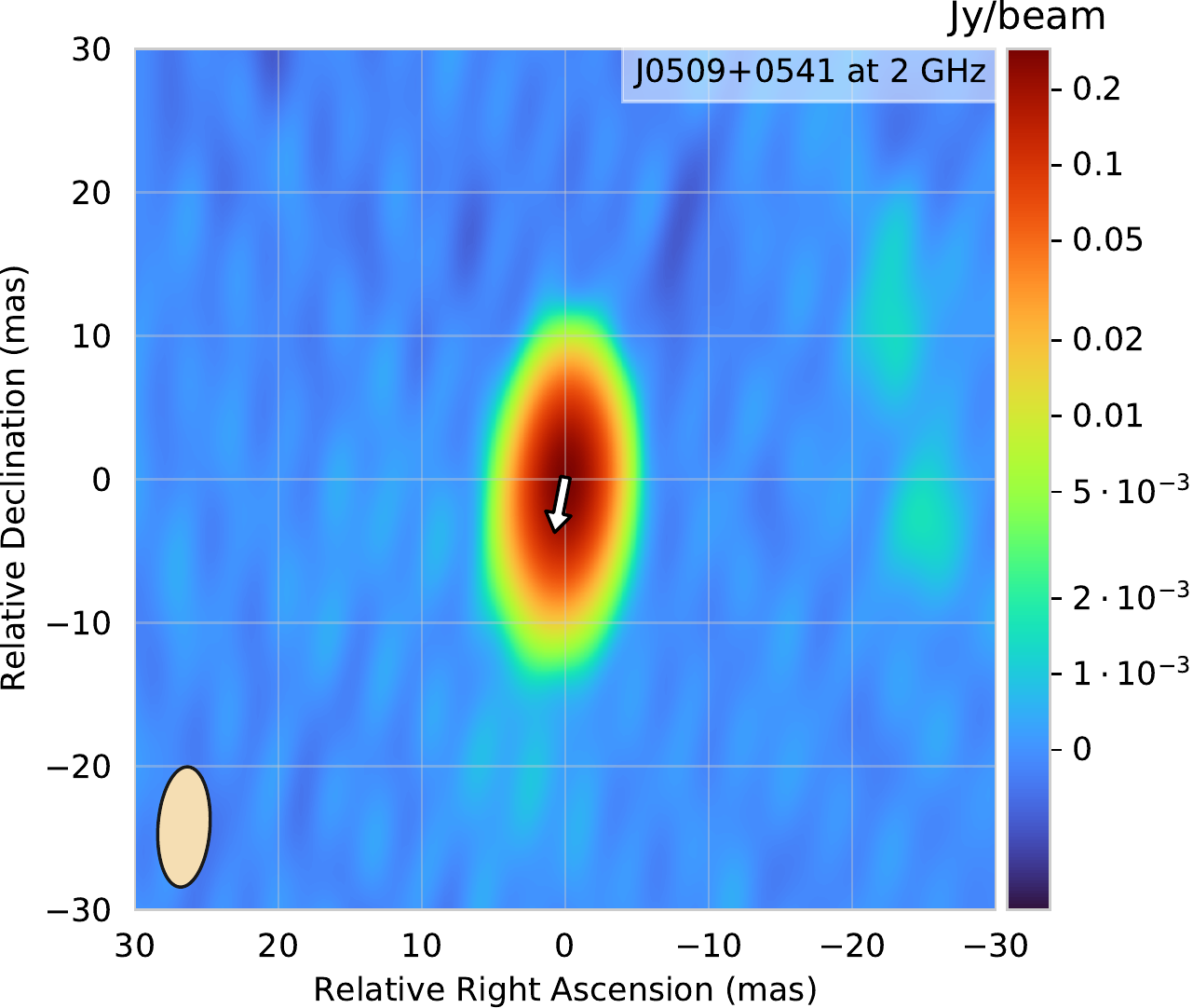}
\includegraphics[height=0.22\textheight]{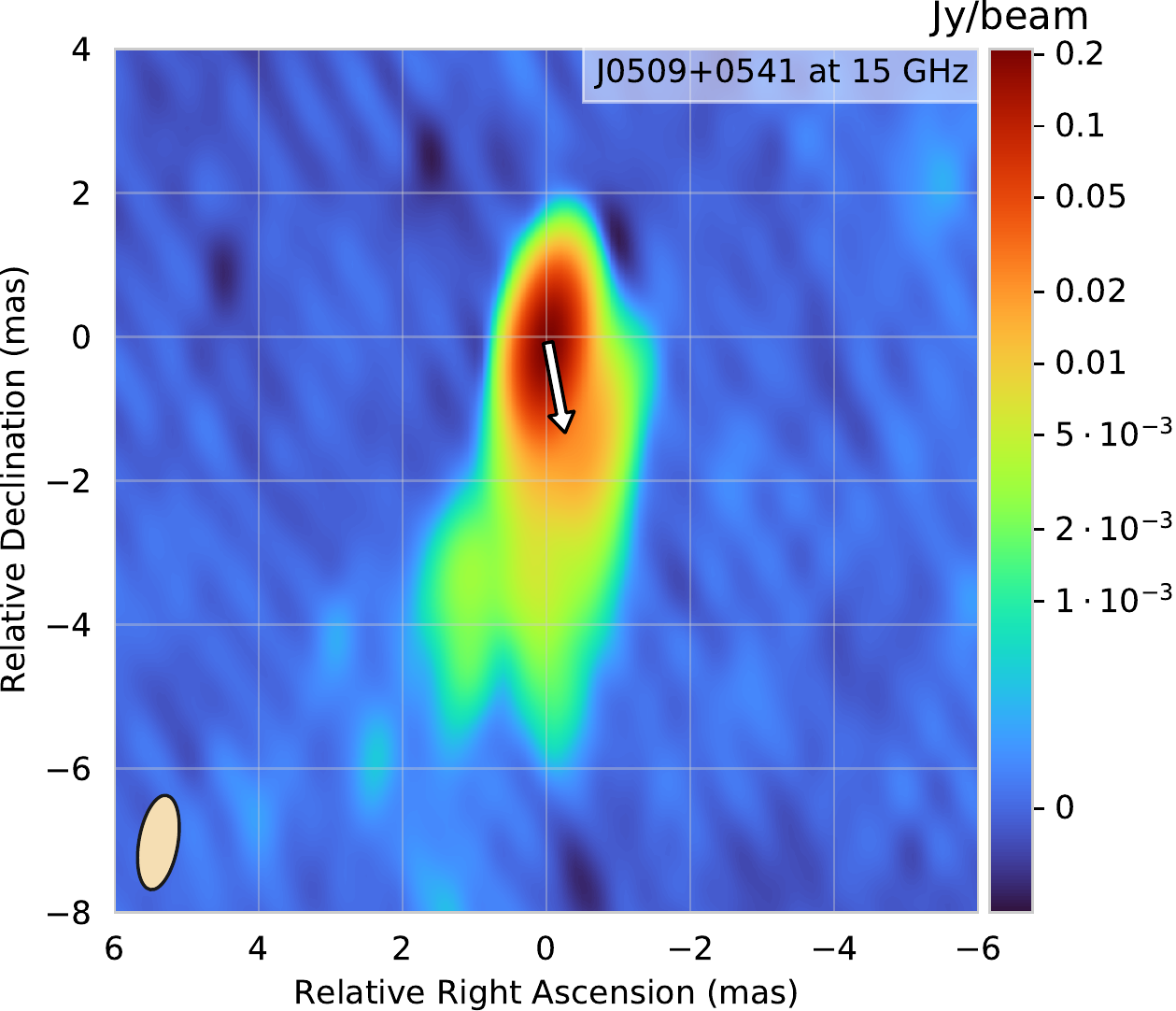}
\caption{VLBI images illustrating the jet direction determination at the shortest angular scales possible. Note the arrow length that corresponds to the separation between the two model components. \emph{Left:} an example of a straight jet. \emph{Middle:} the jet is not visible at the restored image, but our visibility-based approach successfully detects it. See the \emph{right} panel showing the same AGN observed at a higher frequency: the jet orientation is in agreement.}
\label{f:example_short}
\end{figure*}

It is unlikely for a jet component to be stronger and brighter than the core even at the highest available frequency, though it can be possible. In fact, we are aware of two AGNs where this effect was found in dedicated studies: J0900$-$2808 \citep{Kosogorov2021ParsecScale} and J0927+3902 \citep[4C~+39.25, e.g.,][]{1993ApJ...402..160A}. Taking the results of those works into account, we manually flip our estimated jet directions for these two objects. Their images at the highest frequencies are shown in \autoref{f:example_flipped} overlaid with the corrected jet directions.
The $180\degr$ flips are indicated in \autoref{t:jetdirs_perband}.

A key advantage of visibility-based approaches in describing the observed structure is a higher effective resolution, as discussed in \autoref{s:method}. This is illustrated in \autoref{f:example_short} by examples when the jet direction is successfully measured on the shortest scales present in the data. Here, J0541+5312 has its orientation determined closer to the jet origin than possible using the restored image. This orientation stays essentially the same for the more extended jet. The jet of J0509+0541 is hardly visible in \autoref{f:example_short} (middle) at 1.4~GHz: its emission is basically unresolved in the image. Nevertheless, our model-fitting approach detects the second bright component and determines the correct jet orientation, as evidenced by the comparison with a higher-resolution 15~GHz image.

Effects such as temporal variations \citep[e.g.,][]{2013AJ....146..120L} or apparent bending of the jet influence any method of measuring its direction. Such effects likely have a stronger impact on the results of our method compared to approaches based on restored VLBI images or observation at other instruments; this is due to a higher achieved resolution. Indeed, temporal and spatial variations are typically more pronounced closer to the central engine: see \autoref{f:example_diff28} for an example where directions determined at 2 and 8~GHz noticeably differ. This is caused by different spectral properties of emission regions and by VLBI resolution getting proportionally higher with increasing frequency. We stress that both orientation estimates are correct, they effectively probe different angular scales in the jet. We attempt to estimate the innermost jet direction, and they can fundamentally be different at different observing frequencies.

Determining jet directions at lower and higher frequencies has different advantages. We perform the final per-source aggregation to make our results useful for a wider range of studies. That is, we combine and average all measurements for each AGN together across all observing frequencies. These averages are provided in \autoref{t:jetdirs_agg} with uncertainties estimated on the basis of individual errors at each frequency. At the same time, for studies that require probing different spatial scales, we present band-specific directions in \autoref{t:jetdirs_perband}.

\subsection{Statistical Comparisons}

\begin{figure}
\centering
\includegraphics[width=\linewidth,trim=0cm 0cm 0cm -0.3cm]{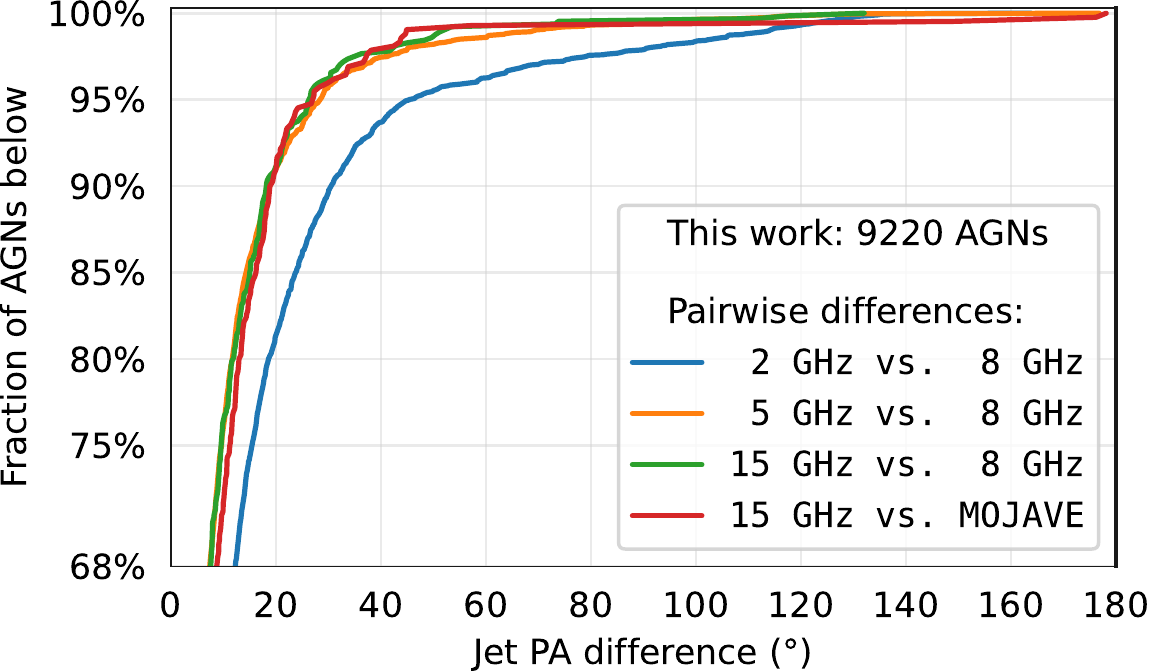}
\caption{Cumulative distribution of differences between jet directions in the same AGN determined at frequency pairs with the largest numbers of overlapping objects: (2--8)~GHz, (5--8)~GHz, (8--15)~GHz. Additionally, we show the comparison of 15~GHz measurements to those estimated within the MOJAVE program \citep[][uses the same 15 GHz observations]{2019ApJ...874...43L}.}
\label{f:pa_diffs}
\end{figure}

\begin{figure}
\centering
\includegraphics[width=\linewidth,trim=0cm 0cm 0cm -0.3cm]{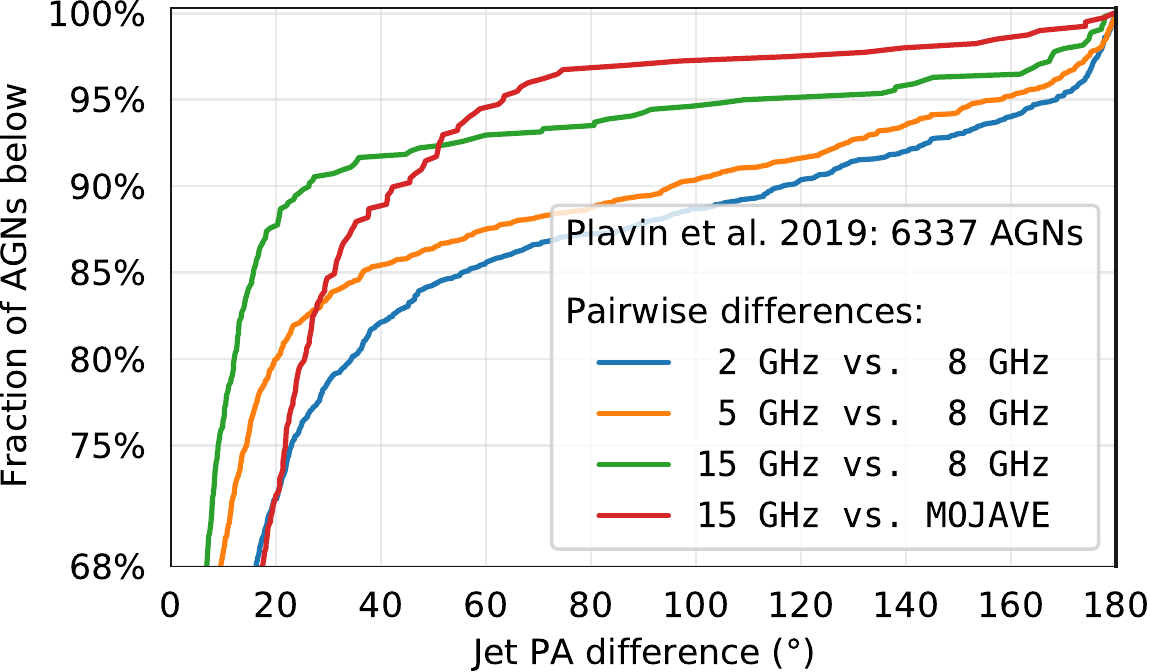}
\includegraphics[width=\linewidth,trim=0cm 0cm 0cm -0.3cm]{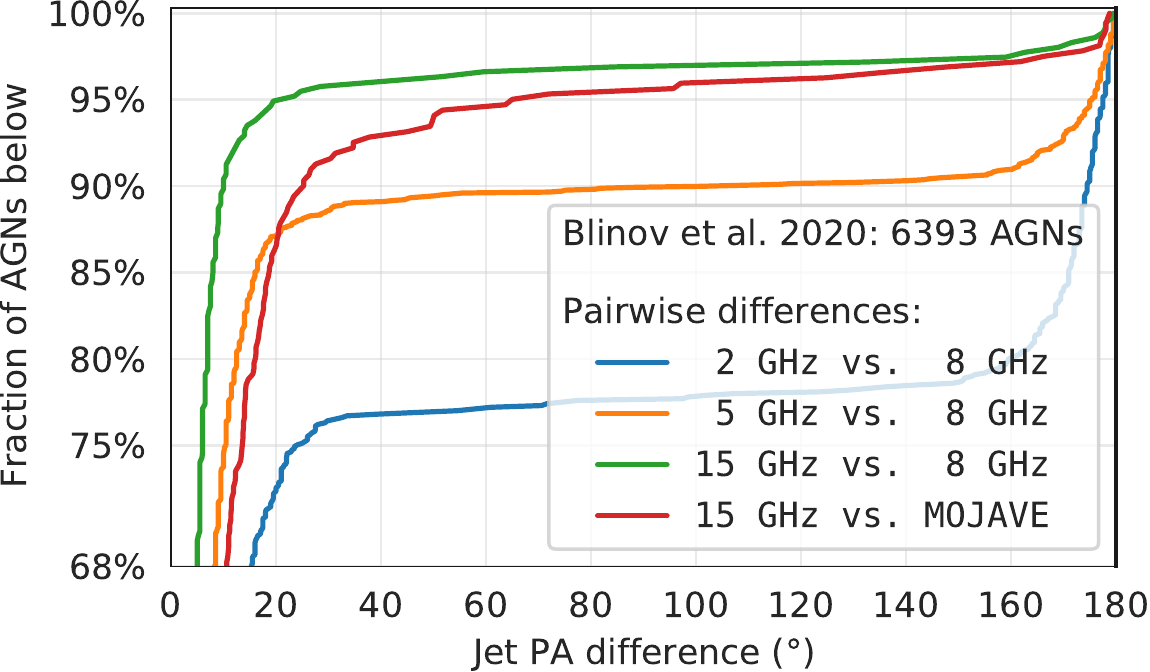}
\caption{Distribution of jet direction differences: same as \autoref{f:pa_diffs}, but for measurements performed in two other works, top: \citet{2019ApJ...871..143P} and bottom: \citet{2020A&A...635A.102B}.}
\label{f:pa_diffs_other}
\end{figure}

We systematically quantify the performance of our jet direction measurements in two ways. First, we check for internal consistency by comparing orientations for the same AGN estimated at different frequencies. Second, we compare our results to the inner jet directions from the MOJAVE program \citep{2019ApJ...874...43L}; we utilize the same calibrated 15~GHz VLBA observations as they do. The directions from \citet{2019ApJ...874...43L} are based on a uniform and detailed manual model-fitting of the jet structure seen by VLBA, and the corresponding models are extensively tested for robustness. The results of both comparisons are presented in \autoref{f:pa_diffs}. Note, that the figure only includes AGNs where the jet orientation was determined; see \autoref{f:images_cnt}. We show the distribution of jet direction differences between different frequency pairs: 2~GHz and 8~GHz, 5~GHz and 8~GHz, 8~GHz and 15~GHz. These are pairs with the largest number of overlapping AGNs. The pairs that do not include the lowest frequency of 2~GHz show an excellent agreement, with the differences less than $10^\circ$ for 68\% of objects and less than $30^\circ$ for 95\%. Measurements at 2~GHz are consistent with higher frequencies but have a somewhat higher spread: up to $45^\circ$ at a 95\% level compared to 8~GHz. Qualitatively, this is the expected behavior due to both the resolution being inversely proportional to frequency, and the difference in core and jet spectra (\autoref{s:method}). \autoref{f:pa_diffs} also indicates that our 15~GHz measurements agree very well with the corresponding MOJAVE ones \citep{2019ApJ...874...43L}. They use the same VLBI observations, a similar general approach, but fit more detailed models that get manually examined to ensure robustness, from which the jet direction was derived based on the feature nearest to the core over all epochs.

Finally, we compare the accuracy of the jet direction measurements presented here with existing results based on restored VLBI images. An earlier work of ours \citep{2019ApJ...871..143P} relied on parsec-scale jet orientations to gain astrophysical insights from a radio-optical comparison. The apparent directions of 6337 AGN jets were estimated, even though only 4023 of those corresponded to the VLBI-Gaia matches and were utilized in that study.
The inter-band differences among these measurements are shown in \autoref{f:pa_diffs_other} and are generally consistent between the frequencies. Comparison with \autoref{f:pa_diffs} clearly shows that in the current work, we significantly improve on both quantity and quality. Indeed, there are more AGNs with direction measurements, which is partially explained by additional VLBI observations that became available. All orientation differences~--- inter-band and relative to MOJAVE~--- decreased, indicating more accurate direction estimates.

A more recent study by \citet{2020A&A...635A.102B} explored the global alignment of parsec-scale AGN jets; they measured jet orientations in a semi-manual way. We show the corresponding diagnostic plots in \autoref{f:pa_diffs_other}. The number of AGNs with orientation estimates remained essentially the same, as in \citet{2019ApJ...871..143P}, and is significantly lower than we present in this work. \autoref{f:pa_diffs_other} shows a good average consistency of direction estimates, as evidenced by 68\% quantiles.
However, the resolution of the $180\degr$ ambiguity was not performed, as these ambiguities did not affect the analysis in \citet{2020A&A...635A.102B}. They lead to larger 95\%-level differences,W especially prominent at lower frequencies, 2~GHz and 5~GHz.
A notable exception is the agreement between 8~GHz and 15~GHz directions: it is tighter for \citet{2020A&A...635A.102B} measurements compared to our results for the majority of objects. We believe this exception is caused by image-based approaches probing somewhat larger angular scales: they can be less affected by spacial and temporal variations (see \autoref{s:method}).

Neither \citet{2019ApJ...871..143P} nor \citet{2020A&A...635A.102B} included tables with jet direction measurements. We obtained the results of the latter in private communication with the authors.

\section{Summary}

This paper presents a novel, completely automatic approach to estimating parsec-scale jet directions based on VLBI observations ranging from 1.4~GHz to 86~GHz. We use visibilities, the primary Fourier-space interferometric observables, to achieve the highest possible effective resolution. We apply this method to calibrated VLBI observations collected in the Astrogeo database and measure positional angles for the largest sample of parsec-scale AGN jets to date, which consists of \jetdirAGNs{} objects. The results are presented in \autoref{t:jetdirs_perband} separately for each available frequency band, and in \autoref{t:jetdirs_agg}, as aggregated averages with a single direction for each AGN. The probed angular scales range from a tenth of a milliarcsecond to tens of milliarcseconds (\autoref{f:corejet_dist}).

We demonstrate the performance of our approach and the robustness of its results by analyzing the consistency of the jet orientation at different dates and different observing frequencies. Further, we find a good agreement with jet directions measured by the MOJAVE team. The automatic method presented in this paper should be used with care when performing detailed studies of individual AGNs: description of the observed emission with two components can be too simplistic for jets with a complex resolved structure; treating the jet orientation as a single number can be suboptimal when variations with time, frequency, or distance are present. Nevertheless, our measurement results are justified on a statistical basis and are useful for studies of large AGN samples. Orientations estimated for each observing frequency separately (\autoref{t:jetdirs_perband}) can further be used as an indicator of variations between spatial scales.

\section*{Acknowledgements}

This research was supported by the Russian Basic Research Foundation grant 19-32-90141.
We thank the teams referred to in \autoref{s:obsdata} for making their fully calibrated VLBI FITS data publicly available, Dmitry Blinov for providing jet direction measurements from \cite{2020A&A...635A.102B}, Leonid Petrov, Richard Porcas, Eduardo Ros and an anonymous referee for helpful discussions and comments.
We are grateful to Elena Bazanova for English language editing and proofreading of the text.
This research has made use of data from the MOJAVE database\footnote{\url{http://www.physics.purdue.edu/astro/MOJAVE/}} that is maintained by the MOJAVE team \citep{2018ApJS..234...12L}.
This study makes use of the VLBA data from the Boston University VLBA Blazar Monitoring Program (BEAM-ME and VLBA-BU-BLAZAR)\footnote{\url{http://www.bu.edu/blazars/BEAM-ME.html}}, funded by NASA through the Fermi Guest Investigator Program.
This research has made use of NASA’s Astrophysics Data System. This research has made use of the NASA/IPAC Extragalactic Database (NED), which is funded by the National Aeronautics and Space Administration and operated by the California Institute of Technology.

\bibliographystyle{aasjournal}
\bibliography{jet_directions}

\end{document}